\documentclass[apjl,a4paper,12pt,useAMS]{emulateapj}
\usepackage{amssymb}
\usepackage{graphicx}

\newcommand{\be}{\begin{eqnarray}}
\newcommand{\ee}{\end{eqnarray}}

\begin{document}
\title{Constraining the age of a magnetar possibly associated with FRB 121102}
\author{Xiao-Feng Cao$^{1}$, Yun-Wei Yu$^{2,3}$, and Zi-Gao Dai$^{4,5}$}
\altaffiltext{1}{School of Physics and Electronics Information,
Hubei University of Education, Wuhan 430205, China}
\altaffiltext{2}{Institute of Astrophysics, Central China Normal
University, Wuhan 430079, China, {yuyw@mail.ccnu.edu.cn}}
\altaffiltext{3}{Key Laboratory of Quark and Lepton Physics (Central
China Normal University), Ministry of Education, Wuhan 430079,
China}
\altaffiltext{4}{School of Astronomy and Space Science, Nanjing University, Nanjing 210093, China}
\altaffiltext{5}{Key Laboratory of Modern Astronomy and Astrophysics (Nanjing University), Ministry of Education, China}

\begin{abstract}
The similarity of the host galaxy of FRB 121102 with those of long
gamma-ray bursts and Type I super-luminous supernovae suggests that
this FRB could be associated with a young magnetar. By assuming the
FRB emission to be produced within the magnetosphere, we derive a lower
limit on the age of the magnetar, after which GHz emission is able to
escape freely from the dense relativistic wind of the magnetar. Another
lower limit is obtained by requiring the dispersion measure
contributed by the electron/positron pair wind to be consistent with
the observations of the host galaxy. Furthermore, we also derive some upper limits on the
magnetar age with discussions on possible energy sources of the FRB emission and the recently-discovered persistent radio counterpart. As a result, some constraints on model parameters are addressed by reconciling the lower limits with the possible upper limits that are derived with an assumption of rotational energy source.
\end{abstract}

\section{Introduction\label{sec:intro}}
Fast radio bursts (FRBs) are mysterious radio transients that have been observed to have
typical durations of a few milliseconds and fluxes of up to a few
tens of Jansky at $\sim1$ GHz (Lorimer et al. 2007; Thornton et al. 2013; Burke-Spolaor \& Bannister 2014;
Spitler et al. 2014; Ravi et al. 2015; Champion
et al. 2016). Although their physical origin is unknown, FRBs are
widely believed to come from cosmological distances in view of their
anomalously high dispersion measures
(DMs; $\sim 300-1600\rm pc~cm^{-3}$) that are difficult to be
accounted for by Galactic high-latitude objects. Thus, the peak radio
luminosity and total energy release of FRBs are estimated to
$\sim10^{42}-10^{43}~\rm erg ~s^{-1}$ and $\sim10^{39}-10^{40}\, \rm erg$,
respectively (e.g. Bera 2016; Cao et al. 2017).

According to the millisecond timescale and high energy release, on
one hand, catastrophic collapses/mergers of compact star systems
are often proposed to be responsible for FRBs, including collapses of supra-massive neutron stars to black holes at
several thousand to million years old (Falcke \& Rezzolla 2014) or
at birth (Zhang 2014), inspiral or mergers of double neutron stars
(Totani 2013; Wang et al. 2016), mergers of binary white dwarfs
(Kashiyama et al. 2013), mergers of charged black holes (Zhang 2016), and collisions
of asteroids/comets with neutron stars (Geng \& Huang 2015).
On the other hand, in view of the coherent
emission property, FRBs are often connected with some energetic
activities of pulsars (more specifically, magnetars), such as
giant flares of soft gamma-ray repeaters (Popov \& Postnov 2010),
synchrotron maser emission from relativistic, magnetized shocks due
to magnetar flares (Lyubarsky 2014), supergiant pulses from
pulsars (Cordes \& Wasserman 2016), encounters of pulsars and asteroid belts (Dai et al. 2016),
accretion onto a neutron star from its magnetized white dwarf companion (Gu et al. 2016),
and pulsars suddenly ``combed" by a nearby
strong plasma stream (Zhang 2017). In addition, some exotic models
have also be proposed such as oscillations of superconducting cosmic
string loops (Cai et al. 2012; Yu et al. 2014).

More observational constraints are undoubtedly necessary for
distinguishing between these models. An important clue has been provided by
the discovery of a repeated FRB in the Arecibo Pulsar ALFA Survey
on 2012 November 2 (Spitler et al. 2014), which was surprisingly
detected again on 2015 May 17 and June 2 (Spitler et al. 2016;
Scholz et al. 2016). These long time gaps (both 572 and 23 days)
make the catastrophic models difficult to be saved. More
excitingly, unambiguous multi-wavelength counterparts of FRB 121102
have been recently captured and identified by Chatterjee et al.
(2017), Marcote et al. (2017), and Tendulkar et al. (2017),
including a persistent radio source and a low-metallicity,
star-forming dwarf galaxy. The detection of the host galaxy helped
to determine the redshift of FRB 121102 to $z=0.19273(8)$, which
corresponds to a luminosity distance of 972 Mpc and undoubtedly
confirms the cosmological origin of the FRB (i.e. only FRB 121102 has been confirmed, not FRBs in general).

The host galaxy of FRB 121102 was found to share many common
properties with those of long gamma-ray bursts (GRBs) and Type I
super-luminous supernovae (SLSNe). Therefore, the possibility that
FRB 121102 is associated with a young magnetar is enhanced
significantly, since both GRBs and SLSNe are widely considered to be
powered by a newly-born rapidly rotating magnetar (Usov 1992; Dai \&
Lu 1998a,b; Zhang \& Meszaros 2001; Woosley 2010; Kasen et al.
2010). Such a magnetar is possibly embedded within a young supernova
remnant, which can trap radio emission at early times and result in
a significant contribution to DM (Kulkarni et al. 2015; Piro 2016;
Murase et al. 2016). Following this idea, the age of the magnetar of FRB 121102 was
constrained to be about a few decades by Metzger et al. (2017).
Nevertheless, such a supernova remnant is not indispensable, even
for the explanation of the observed persistent radio emission (Dai
et al. 2017).

In any case, before the FRB emission encounters the supernova
ejecta, it probably has to first penetrate a dense relativistic wind
from the magnetar, if the emission is produced within the stellar
magnetosphere. As first suggested by Yu (2014), this pulsar wind,
which consists of an extremely great number of electrons and
positrons, can cause a more serious radio trapping and a more
significant DM contribution than the supernova ejecta, in
particular, by considering of the relativistic boosting effect. In
this Letter, therefore, we derive a more stringent constraint on the
age of the magnetar of FRB 121102, by constricting the
electron/positron loading of the magnetar wind to be consistent with
the implications from the host galaxy observations.

\section{Lower limits on spin periods and magnetar ages}
By considering of a magnetar of an angular frequency, $\Omega$, a surface polar
strength of magnetic field, $B_{\rm p}$, and a radius, $R=10^6$ cm, the
electron-positron distribution of the magnetar magnetosphere can be described as usual by the Goldreich \& Julian (GJ)
particle density $n_{\rm GJ}(r)\approx({\Omega B_{\rm p}/ 2\pi c
e})({r/ R})^{-3}$ (Goldreich \& Julian 1969), the
angle-dependence of which is ignored for a simple order-of-magnitude analysis. Beyond the light cylindrical radius $r_{\rm
L}=c/\Omega$, the corotation of the magnetosphere can no longer be
held. The magnetocentrifugal force exerting on the plasma and
especially the subsequent magnetic reconnections will launch a
relativistic wind. The particle number flux of the wind can be expressed as $\dot{N}_{\rm w}\approx4\pi r_{\rm L}^2\mu_{\pm} n_{\rm GJ}(r_{\rm
L})c$, where $P=2\pi/\Omega$ is the spin period. The $e^{\pm}$ multiplicity parameter, $\mu_{\pm}$, represents a ratio of the wind flux to the GJ flux, because a great number of electrons and positrons could be
generated spontaneously as the wind propagation and energy
dissipation. Then the density of the wind at radius $r$ can be expressed as
\begin{eqnarray}
n_{\rm w}(r)\approx{\dot{N}_{\rm w}\over 4\pi r^2 c }={\mu_{\pm}
n_{\rm GJ}(r_{\rm L})}\left({r\over r_{\rm L}}\right)^{-2}.
\end{eqnarray}

On one hand, with the above density, the plasma frequency of the magnetar wind at
different radii can be calculated by
\begin{eqnarray}
{\nu_{\rm p}}(r)={\Gamma\over1+z}\left[{e^2\over\pi m_{\rm
e}}{\mu_{\pm} n_{\rm GJ}(r_{\rm L})\over
\Gamma}\right]^{1/2}\left({r\over r_{\rm L}}\right)^{-1}
\end{eqnarray}
for $r\geq r_{\rm L}$, where $\Gamma$ is the Lorentz factor of the wind at the radius.
Following Drenkhahn (2002), a reference dynamical result can be adopted as $\Gamma=\Gamma_{\rm L}(r/r_{\rm L})^{1/3}$ and then we have
\begin{eqnarray}
{\nu_{\rm p}}(r)={\mu_{\pm}^{1/2}\Gamma_{\rm L}^{1/2}\over1+z}\left[{e^2\over\pi
m_{\rm e}}{n_{\rm GJ}(r_{\rm L})}\right]^{1/2}\left({r\over r_{\rm
L}}\right)^{-5/6}.
\end{eqnarray}
The initial velocity of the wind can be set to the Alfv{\'e}n
velocity at the light cylinder as $\Gamma_{\rm
L}\sim\sqrt{\sigma_{\rm L}}$, where $\sigma_{\rm L}$ represents the
initial ratio between Poynting flux to matter energy flux.
Since the total energy flux carried by the magnetar wind is
completely provided by the magnetic spin-down of the magnetar, i.e.,
$(\sigma_{\rm L}+1)\Gamma_{\rm L}\dot{N}_{\rm w}m_{\rm e}c^2=L_{\rm
sd}$, the Lorentz factor at the light cylinder can be determined
by $\Gamma_{\rm L}\sim({L_{\rm sd}/ \dot{N}_{\rm w}m_{\rm
e}c^2})^{1/3}$. By taking the spin-down luminosity as usual as $L_{\rm sd}={B_{\rm p}^2R^6\Omega^4/( 6c^3)}$, the maximum value of plasma frequencies of the magnetar wind can be written as
\begin{eqnarray}
{\nu_{\rm p}}(r_{\rm L})=1.5\times10^4\mu_{\pm}^{1/3}B_{\rm p,
14}^{2/3}P_{-3}^{-7/3}\rm GHz,
\end{eqnarray}
outside of the light cylinder. Hereafter the conventional notation $Q_{x}=Q/10^x$ is adopted in cgs units. In order to guarantee that GHz
radio emission can freely penetrate the wind, the above plasma
frequency should not be higher than the radio frequency and thus we
can obtain the spin period as
\begin{eqnarray}
P>61\mu_{\pm}^{1/7}B_{\rm p, 14}^{2/7}\rm ms.
\end{eqnarray}
This lower limit is much longer than the initial spin period of a few
milliseconds that can be inferred from SLSN or GRB observations.
Therefore the corresponding age of the magnetar can be constrained to
\begin{eqnarray}
t_{\rm age}>24\mu_{\pm}^{2/7}B_{\rm p, 14}^{-10/7}\rm
yr,\label{t_mw_pf}
\end{eqnarray}
which is calculated with the expression of spin-down timescale as
$t_{\rm sd}=3Ic^3/(B_{\rm p}^2R^6\Omega^2)$, where $I=10^{45}~\rm g~cm^2$ is the moment of inertial of the magnetar. Correspondingly, the age constraint due to the radio trapping of supernova ejecta can be derived from Equation (9) of Metzger et al. (2017) to
\begin{eqnarray}
t_{\rm age}>0.7f_{\rm ion,-1}^{1/3}\left({M_{\rm ej}\over 10M_{\odot}}\right)^{1/3}v_{\rm ej,9}^{-1}\rm ~ yr,\label{t_sn_pf}
\end{eqnarray}
where $f_{\rm ion}$, $M_{\rm ej}$, and $v_{\rm ej}$ are the ionization fraction, the mass, and the velocity of the ejecta. Obviously, the constraint presented in Equation (\ref{t_sn_pf}) can be ignored safely in comparison with the more stringent constraint given by the wind plasma frequency as shown in Equation (\ref{t_mw_pf}).

%
\begin{figure*}
\centering\resizebox{0.3\textwidth}{!} {\includegraphics{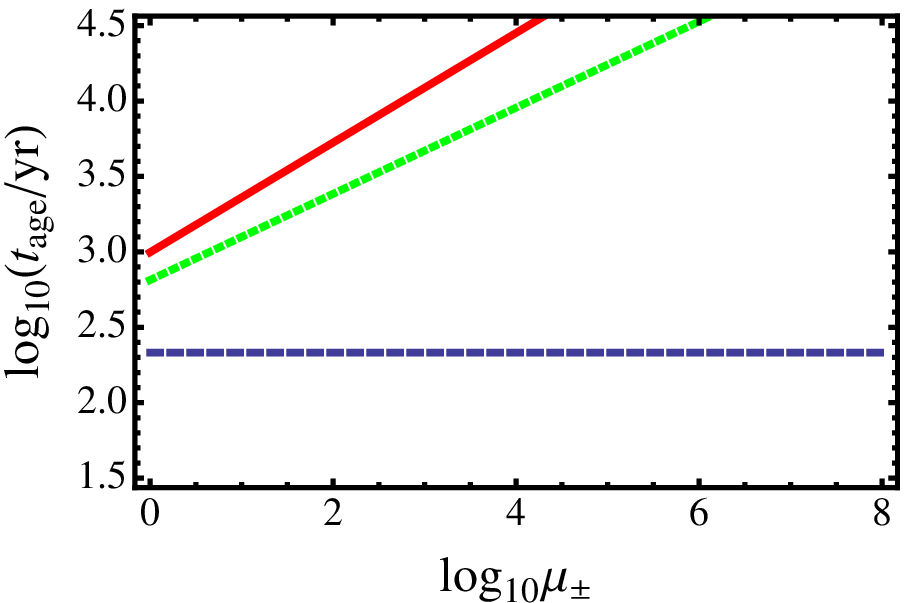}}\resizebox{0.3\textwidth}{!} {\includegraphics{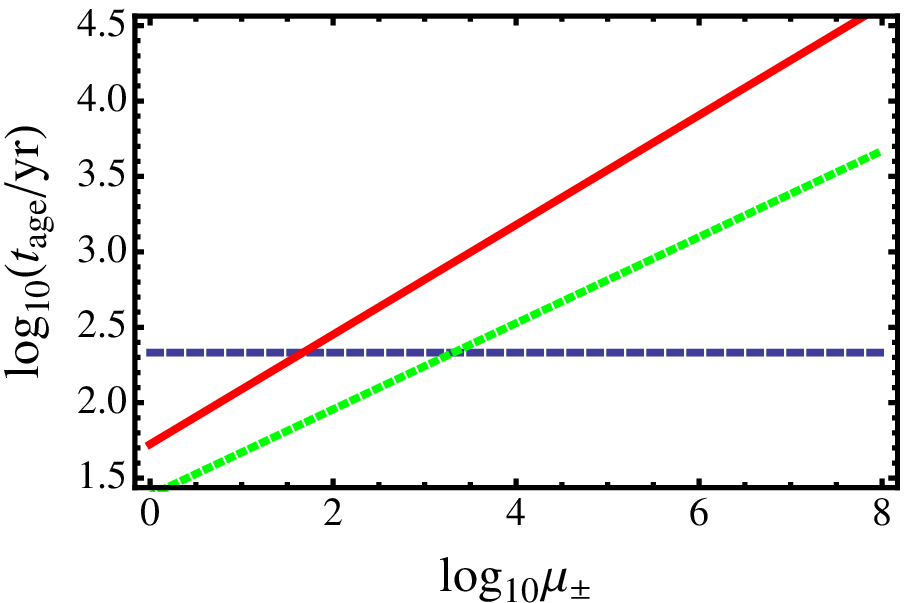}}\resizebox{0.3\textwidth}{!} {\includegraphics{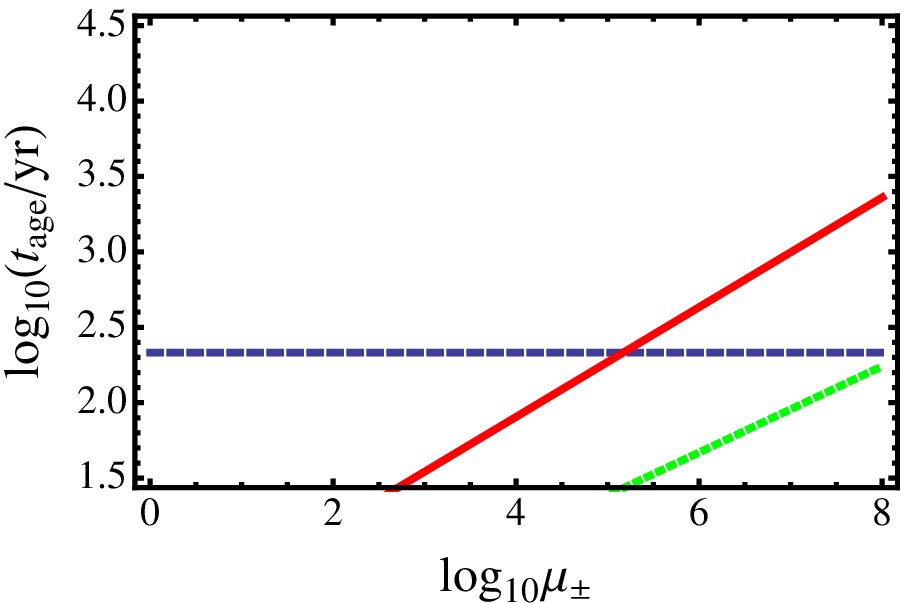}}
\caption{Lower limits on the magnetar age that are obtained by constraining the wind plasma frequency (dotted green line), the wind DM contribution (solid red line), and the DM contribution of supernova ejecta (dashed blue line), where a putative upper limit on the DM of the FRB source is taken as $\rm DM_{src,up}=1pc~cm^{-3}$. The panels from left to right correspond to magnetic fields of $10^{13}$ G, $10^{14}$ G, and $10^{15}$ G, respectively.}
\end{figure*}
%

On the other hand, by taking the
relativistic boosting effect into account, the DM contributed by the magnetar wind can
be calculated by (Yu 2014)
\begin{eqnarray}
{\rm DM}_{\rm w}&=&{1\over (1+z) }\int_{r_{\rm L}} 2\Gamma(r)\cdot
n_{\rm w}(r) dr\nonumber\\
&=&{3\Gamma_{\rm L}\mu_{\pm}n_{\rm GJ}(r_{\rm L})r_{\rm L}\over (1+z)}\nonumber\\
&=&1.5\times10^{7}\mu_{\pm}^{2/3}B_{\rm
p,14}^{4/3}P_{-3}^{-11/3}~\rm pc~cm^{-3}.\label{DMnsw}
\end{eqnarray}
For an upper limit value of DM$_{\rm w,up}$, the spin period of the magnetar can be constrained to
\begin{eqnarray}
P>90\mu_{\pm}^{2/11}B_{\rm p,14}^{4/11}{\rm
DM^{-3/11}_{w,up}}\rm ms,
\end{eqnarray}
which corresponds to an age of
\begin{eqnarray}
t_{\rm age}>53\mu_{\pm}^{4/11}B_{\rm p,14}^{-14/11}\rm DM^{-6/11}_{w,up}
yr.\label{t_low_w}
\end{eqnarray}
For a comparison, the age constraint from the DM constraint on the supernova ejecta can be written as follows:
\begin{eqnarray}
t_{\rm age}>215f_{\rm ion,-1}^{1/2}\left({M_{\rm ej}\over 10M_{\odot}}\right)^{1/2}v_{\rm ej,9}^{-1}\rm DM_{ej, up}^{-1/2}~ yr,\label{t_low_sn}
\end{eqnarray}
where $\rm DM_{ej, up}$ is the upper limit on the DM of the ejecta. This expression is derived from Equation (12) of Metzger et al. (2017), which was however not addressed there because they did not separated the DM of the FRB source from that of the host galaxy. By comparing Equation (\ref{t_low_w}) with (\ref{t_low_sn}), the age constraint given by the magnetar wind can be more stringent than the supernova ejecta constraint as long as $\mu_{\pm}^{4/11}B_{\rm p,14}^{-14/11}>5$.

\section{Discussions on FRB 121102}
For FRB 121102, its total DM was measured to be $\rm DM_{\rm
total}=558\rm ~pc~cm^{-3}$ (Spitler et al. 2016; Chatterjee et al.
2017), which is contributed jointly by the Milky Way and its halo,
the intergalactic medium, the host galaxy, and the FRB source
itself. As analyzed by Tendulkar et al. (2017), the sum of the last
two contributions can be constrained to $55\lesssim \rm
(DM_{host}+DM_{src})\lesssim 225pc~cm^{-3}$. Moreover, according to
Equation 6 of Tendulkar et al. (2017), the value of $\rm DM_{host}$
is probably not much lower than $\sim 100\rm ~pc~cm^{-3}$, if FRB
121102 does not offset very much from its host galaxy. Therefore,
the DM contribution leaving to the FRB source including the magnetar
wind and the supernova remnant is very limited, which is consistent
with the small fluctuation of the DM of FRB 121102 during the past
few years (Spitler et al. 2016). For a putative DM upper limit of
$\rm DM_{src,up}=1pc~cm^{-3}$ and three typical magnetic fields, the
different lower limits on the magnetar age are presented in Figure
1, as functions of the uncertain $e^{\pm}$ multiplicity. According
to observations of pulsar wind nebulae, $\mu_{\pm}$ is usually
considered to be very high, since a great number of electrons and
positrons are needed to produce the observed wind emission and to
determine a typical Lorentz factor of $\sim10^4-10^5$ of the wind
(e.g. Yu et al. 2014).

As a general result, the lower limit on the
magnetar age can be found to be, at least, about a few hundreds to thousands
years. Then, a question could arise: whether or not the rotational energy of the magnetar can power the FRB emission and also the persistent radio counterpart. First of all, according to some observations of Galactic pulsars, it has been suggested that FRBs could be analogical to giant pulses that are powered by the spin-down of a magnetar. In this case, however, the magnetar age of FRB 121102 would be constrained to be $t_{\rm age}<9(L_{\rm FRB}/10^{41}{\rm erg~s^{-1}})^{-1}B_{\rm p,14}^{-1}$ yr (Metzger et al. 2017), which is probably in conflict with the lower limits given above. Moreover, observationally, giant pulses from the Crab pulsar, statistically, only have an energy release $\Delta E\sim 10^{28}\,$erg per giant pulse (Majid et al. 2011), which indicates a ratio of this energy to the total stellar rotational energy as $\Delta E/E_{\rm rot}<5\times10^{-22}$. If a same emission mechanism is assumed for FRB 121102, then an absolutely impossible rotational energy of $\sim 10^{60}$ erg would be required to explain the isotropic-equivalent FRB energy of $\Delta E\sim E_{\rm iso} =2\times10^{39}$ erg. This difficulty was also recently pointed out by Lyutikov (2017).

As an alternative scenario, we here propose that the energy release
of an FRB is connected with a glitch-like process, although the
physics of this process is completely unknown. By denoting the
sudden change of spin frequency by $\Delta \Omega$, the energy
release can be calculated by
\begin{eqnarray}
\Delta E={1\over2}I\Omega^2-{1\over2}I(\Omega-\Delta \Omega)^2\thickapprox I\Omega\Delta \Omega.
\end{eqnarray}
Observations of Galactic pulsars usually found
$\Delta \Omega/\Omega\sim 10^{-9}-10^{-6}$ for their glitches and
the current maximum value can be as large as
$10^{-5}$ (Yuan et al. 2010; Manchester \& Hobbs 2011). Therefore,
for a released energy of $\sim2\times10^{39}$ erg, the
total rotational energy of the magnetar can be constrained to be $E_{\rm
rot}\gg (10^{5}-10^{9})E_{\rm iso}$, where the symbol ``$\gg$" is
used because of the repeatability of FRB 121102. As a result, the
spin period and age of the magnetar can be derived to be $P\ll
\rm (0.1-10)$ s and $t_{\rm age}\ll (67-6.7\times10^5)B_{\rm
p,14}^{-2}\rm yr$. Such upper limits on the magnetar age can in principle be consistent with the obtained lower limits, if FRBs can indeed be associated with the most giant glitches.

In any case, besides the rotational energy, some other energy sources could still be available to power FRBs. The most popular choice could be the magnetic energy within a magnetar, which is of the order $E_{\rm B}\sim 3\times10^{49}B_{\rm int,16}^2$ erg for an internal magnetic field strength of $B_{\rm int}\sim 10^{16}$ G. However, the disadvantage of this model is that no bright radio pulse was detected from the
giant flare of  SGR 1806-20 (Tendulkar et al. 2017). As another possible solution, Dai et al. (2016) proposed that the FRB energy could be provided by the gravitational energy of an asteroid as it is captured by and collides with the magnetar. Such a process can repeat naturally if an asteroid belt is around the magnetar.
In any case, both of the above alternative scenarios can survive from the constraints on the magnetar age, with at least a few hundreds to thousands years.

For the steady radio emission associated with FRB 121102, it was
currently considered to be produced by synchrotron emission of
pulsar wind nebulae (Kashiyama \& Murase 2017; Metzger et al. 2017;
Dai et al. 2017).  By considering that the luminosity of the wind
emission is ultimately determined by the spin-down luminosity of the
magnetar, it is convenient to simply require the spin-down
luminosity to be higher than the luminosity of the steady radio
emission, i.e., $L_{\rm sd}>L_{\rm radio}=3\times10^{38}\rm
erg~s^{-1}$. This gives a very stringent constraint of
\begin{eqnarray}
P<134B_{\rm p,14}^{1/2}\rm~ ms
\end{eqnarray}
and
\begin{eqnarray}
t_{\rm age}<116B_{\rm p,14}^{-1}\rm yr.\label{t_up}
\end{eqnarray}
In any case, if this steady radio emission is not powered by the
spin-down of magnetar, the above constraint can be removed but then
some alternative scenarios should be suggested. By comparing
Equation (\ref{t_up}) with (\ref{t_low_w}) and (\ref{t_low_sn}),
some relationships between the model parameters can be derived as
\begin{eqnarray}
B_{\rm p,14}>0.06\mu_\pm^{4/3}\rm DM_{w,up}^{-2}\label{ieq1}
\end{eqnarray}
and
\begin{eqnarray}
B_{\rm p,14}<0.5\rm DM_{\rm ej,up}^{1/2}\label{ieq2}
\end{eqnarray}
in order to make the lower and upper limits of the magnetar age
consistent with each other. As shown, relatively low magnetic fields
are favored, which indicates FRBs more probably associated with
SLSNe than long GRBs. The coexistence of Equations (\ref{ieq1}) and
(\ref{ieq2})  can further give
\begin{eqnarray}
\mu_{\pm}<5.5\rm DM_{w,up}^{3/2}DM_{ ej,up}^{3/8}.\label{ieq3}
\end{eqnarray}

The value of $\mu_{\pm}$ at radii much
far away from the light cylinder can in principle been inferred from the emission property of the wind. 
So far no hard X-ray/soft gamma-ray counterparts
have been detected within $\sim 1^{\circ}$ of FRB 121102's position
(Scholz et al. 2016; see Petroff et al. 2015 for other FRBs), which could be a nature result of the small value of $\mu_{\pm}$. However, as suggested by Kashiyama \& Murase (2017), Metzger
et al. (2017), and Dai et al. (2017), different from the typical Crab-like nebulae, the wind emission of FRB 121102 could actually
be mainly in the radio band exhibiting as the persistent radio counterpart. In this case, the value of $\mu_{\pm}$ at emitting radii is required to be very high (see Dai et al. 2017 for an estimate of the number of emitting electrons/positrons), which is very much higher than that presented in Equation (\ref{ieq3}) for the
light cylinder radius. It is therefore indicated that the lepton load of magnetar wind could significantly evolve at large radii.

\section{Summary}

The recent discovery of the host galaxy of FRB 121102 implies a
possible connection between FRBs and long GRBs/SLSNe. Combining with
the repeatability of FRB 121102, it was suggested that this FRB
could be associated with a young magnetar and originate from some
activities of the magnetar. In order to test this possibility, we
investigated the important influences on FRB emission from the wind
of magnetar, if the FRB emission is produced in the inner
magnetosphere. Specifically, by evaluating the radio trapping and
the DM contribution by wind electrons/positrons, we derived some
lower limits on the spin periods and ages of magnetars visible as
FRBs and applied these results to the case of FRB 121102. Meanwhile,
some possible upper limits are also discussed by considering that
FRB 121102 and moreover its persistent radio emission could be
powered by the rotational energy of magnetar. By reconciling the
lower and upper limits, some constraints on the model parameters
were revealed. For example, for a putative $\rm DM_{\rm src,up}\sim
5~pc~cm^{-3}$, $\mu_{\pm}\sim100$, and a relatively low magnetic
field of $B_{\rm p}\sim 10^{14}$ G, all of the limits on the age can
reach a consensus at the age of about $\sim100$ yr. According to the
most allowable values of magnetic field strengths, FRBs are
suggested to be more probably associated with SLSNe than long GRBs.

\acknowledgements
We thank Kohta Murase and  Bing Zhang for their comments and X. Zhou for her discussions on pulsar glitches.
This work was supported by the National Basic Research Program (``973'' Program) of China (grant No. 2014CB845800) and the National Natural Science Foundation of China (grant Nos. 11473008 and 11573014). XFC was supported by the National Natural Science Foundation of China (grant Nos. 11303010 and 11673008).

\end{document}